\documentstyle[12pt,epsf]{article}
\begin{document}
\title{NEUTRINOLESS DOUBLE BETA DECAY IN HEAVY DEFORMED NUCLEI\footnote{Work
supported in part by Conacyt under contract 400340-5-3513E}}
\author{Jorge G. Hirsch\\
Departamento de F\'{\i}sica,\\ Centro de Investigaci\'on y de Estudios
Avanzados del I. P. N.,\\ A. P. 14-740 M\'exico 07000 D.F.\\
\and
O. Casta\~nos and P. O. Hess\\
Instituto de Ciencias Nucleares,\\ Universidad Nacional Aut\'onoma de
M\'exico,\\ A. P. 70-543 M\'exico 04510 D.F. }
\date{}
\maketitle
\abstract{The zero neutrino mode of the double beta decay in heavy
deformed nuclei is investigated in the framework of the pseudo SU(3)
model, which has provided an accurate description of collective nuclear
structure and predicted
half-lives for the two neutrino mode in good agreement with experiments.
In the case of $^{238}U$ the calculated zero neutrino half-life is at least
three orders of
magnitude greater than the two neutrino one, giving strong support of the
identification of the radiochemically determined half-life as being the two
neutrino double beta decay. For $^{150}Nd$ the zero neutrino matrix
elements are of the
order of magnitude of, but lesser than, those evaluated using the QRPA.
This result confirms that different nuclear models produce similar zero
neutrino matrix elements, contrary to the two neutrino case.
Using these pseudo SU(3) results  and the upper limit for
the neutrino mass we estimate the $\beta\beta_{0\nu}$ half-lives for six
nuclei. An upper limit for majoron coupling constant is extracted from the
experimental data.}

\section{Introduction}

\vskip 1.0pc

Many grand unified theories \cite{Ver86,Fae88,Tom91} regard neutrinos as
Majorana
particles, identical to its own antiparticle, predicting light neutrino
masses and the existence of right-handed currents. In order to get
information about this physics beyond the standard model many extensive
experimental investigations have been performed to detect the
neutrinoless mode of the double beta decay ($\beta\beta_{0\nu}$).

Double beta decay can be classified into various modes according to the
light particles besides the electrons associated with the decay. The two
neutrino mode ($\beta \beta_{2\nu}$), in which two electrons and two neutrinos
are emitted, takes place independently of the neutrino properties, and
it is the only one which has been measured up to now
\cite{Ell87,Kaw93,Ber92,Moe91,Mil90,Boe91,Tur91}.

The nonconserving lepton number zero neutrino mode ($\beta \beta_{0\nu}$),
in which only two electrons are emitted, can occur if, and only if, the
neutrino is a massive Majorana particle \cite{Doi85}, {\it i.e.} if $\nu$
and $\bar{\nu}$
are just the left and right handed helicity states of a single particle
state $\nu^M$. For this reason this mode, unobserved up to now, concentrates
the interest concerned with new physics, beyond the standard model.

Another proposed double beta decay mode is the decay in two electrons and
a massless neutral pseudoscalar $\chi$, known as the majoron. If this
$\beta\beta_{0\nu \chi}$ decay competes with the $\beta\beta_{2\nu }$
mode, the electron energy spectra of the two emitted electrons must be
compared in detail in order to distinguish both decay modes.

In previous papers \cite{Cas94} we used the pseudo SU(3) shell model to
evaluate the
two neutrino double beta half lifes of eleven heavy deformed potential
double beta emitters. We found good agreement with the available
experimental information. The radiochemically measured $^{238}U$ decay
\cite{Tur91} raised some expectatives, given the experimental array cannot
discriminate between the different double beta decay modes, and some
theoretical estimations\cite{Sta90} predicted similar decay ratios for both
the
$0\nu$ and the $2\nu$ modes. Our calculations for the two neutrino mode
were consistent with the experimental result.
In the present work we estimate the zero neutrino matrix elements for six
heavy deformed double beta emitters, including $^{238}U$
and $^{150}Nd$.
In the case of $^{238}U$ we found the zero neutrino half-life at least three
orders of
magnitude greater than the two neutrino one, giving strong support of the
identification of the observed half-life as being the two
neutrino double beta decay.
Experiments in $^{150}Nd$ have set the lowest
limits to
the majoron coupling constant. For $^{150}Nd$ the $\beta\beta_{0\nu}$ matrix
elements are of the
order of magnitude of, but lesser than, those evaluated using the QRPA.
This result confirms the experience that different nuclear models produce
similar zero neutrino matrix elements, contrary to the two neutrino case
\cite{Fae88,Tom91,Sta90,Tom87,Mut89}. Using these pseudo SU(3) results and
the upper limit for
the neutrino mass we estimate the $\beta\beta_{0\nu}$ half-lives for six
nuclei. An upper limit for the majoron coupling constant is also extracted
from the experimental data.

In section 2 we review the double beta decay formalism for the zero
neutrino and majoron emitting modes.  A summary of the pseudo $SU(3)$ model is
given in
Section 3 and in Section 4 the calculations of the nuclear matrix elements
$M_{0\nu} $ within this model is presented.  Section 5
contains the results and discussions and Section 6 the conclusions.
In Appendix A some details of the evaluation of the radial integrals are
given,  the connection with the $2\nu$ formalism is presented in Appendix B.
The inclusion of short range correlations
and the effects related with the finite nucleon size are discussed in
Appendix C.

\vskip 1.0pc
\section {Double beta decay}
\vskip 1.0pc

In the case the $0\nu$ decay exists, the virtual neutrino must be
emitted in
one vertex, and absorbed in the other. Since in the standard theory the
emitted particle is a right-handed antineutrino and the absorbed one a
left-handed neutrino the process requires that a) the exchanged neutrino
is a Majorana particle and b) both neutrinos have a common helicity
component.
The helicity matching can be satisfied in two ways: a) the neutrinos have
a nonvanishing mass and therefore a ``wrong'' helicity component
proportional to $m_\nu / E_\nu$. The decay rate will be proportional to
$<m_\nu >^2$. Or b) the helicity restriction could be satisfied if there
is a right handed current interaction. In this case a nonvanishing
mass allowing mixing of neutrino types is also required \cite{Doi85,Moe94}.

For massive Majorana neutrinos one can perform the integration over the
four-momentum of the exchanged particle and obtain a ``neutrino
potential'' which for a light neutrino ($m_\nu < 10 MeV$) has the form

\begin{equation}
H(r,\overline E) = {\frac {2R}{\pi r}} \int_0^\infty dq {\frac {sin(qr)}
{q+\overline E} }
\end{equation}

\noindent
where $\overline E$ is the average excitation energy of
the intermediate odd-odd nucleus and the nuclear radius $R$ has been
added to make the neutrino potential dimensionless. In the zero neutrino
case this closure approximation is well justified \cite{Pan92}.
The final formula, restricted to the term proportional to the neutrino
mass, is \cite{Tom91,Doi85}
\begin{equation}
(\tau^{1/2}_{0\nu})^{-1} = \left ( {\frac {<m_\nu >}{m_e}} \right )^2
G_{0\nu}  M_{0\nu}^2    .
\end{equation}

\noindent
where $G_{0\nu}$ is the phase space integral associated with the emission
of the two electrons and the nuclear matrix elements
$M_{0\nu}$ are discussed below.

The double beta decay with majoron emission has been investigated
in various models\cite{Tom91,Doi85}. The original Gelmini-Roncadelli
one\cite{Gel81} is incompatible
with the LEP measurement of the invisible width of the Z boson. However,
various suitable modifications have been considered\cite{Moe94,Bere92}. The
decay rate for
the majoron emitting mode of the double beta decay resembles the $0\nu$ mode:

\begin{equation}
(\tau^{1/2}_{0\nu ,\chi })^{-1} = <g_{\nu ,\chi}>^2 G_{0\nu ,\chi}
 M_{0\nu}^2    .
\end{equation}

\noindent
where $<g_{\nu ,\chi}>$ is the effective majoron neutrino coupling constant,
$G_{0\nu .\chi}$ is the phase space integral which describes two
electrons plus the massless majoron, and the nuclear matrix elements
are {\em the same} in both cases.

These nuclear matrix elements are \cite{Doi85}

\begin{equation}
M_{0\nu} \equiv | M_{0\nu}^{GT} - {\frac{g_V^2}{g_A^2}} M_{0\nu}^{F} |
\end{equation}
\noindent
with

\begin{equation}
M_{0\nu}^\alpha =  < 0^+_f \| O^\alpha \| 0^+_i>
\end{equation}
\noindent where the kets
$ \vert 0^+_i \rangle$  and $|0^+_f\rangle$ denote
the corresponding initial and final nuclear  states, the quantities $g_V$
and $g_A$ are the dimensionless coupling constants of the vector and
axial vector nuclear currents, and

\begin{equation}
\begin{array}{l}
O^{GT} \equiv \sum\limits_{m,n} O^{GT}_{mn} = \sum\limits_{m,n}
\vec\sigma_m
t^-_m \cdot \vec\sigma_n t^-_n H(|\vec r_m - \vec r_n |,\overline E) \\
O^{F} \equiv \sum\limits_{m,n} O^{F}_{mn} = \sum\limits_{m,n}  t^-_m t^-_n
H(|\vec r_m - \vec r_n |,\overline E)  .
\end{array}
\end{equation}

\noindent being $\vec\sigma$ the Pauli matrices related with the spin
operator and $t^-$ the isospin lowering operator, which satisfies
$t^-|n> = |p>$. The superindex GT denotes the Gamow-Teller spin-isospin
transfer channel, while the F indicates the Fermi isospin one.

It must be noted that only the difference between Fermi  and
Gamow-Teller  matrix elements appears in the combination $M_{0\nu}$, which
can be evaluated from the new operator

\begin{equation}
O = \sum_{m,n} t^-_m t^-_n H(|\vec r_m - \vec r_n |,\overline E)
( \vec\sigma_m \cdot \vec\sigma_n - ({\frac {g_V} {g_A} } )^2 )
\end{equation}

Although $({\frac {g_A} {g_V}})^2 = 1.5$ for free nucleons
\cite{Sta90,Tom87}, it is common to assume
$({\frac {g_A} {g_V}})^2 = 1.0$ in order to improve the agreement with
the observed Gamow-Teller transition strength, which is quenched by
roughly a factor one half in comparison with the theoretical sum rule
\cite{Vog86}.
In this way, channels which are out of the most usual models, as the
delta-isobar-hole excitations, are included in an effective form
\cite{Hir90}. Under this assumption

\begin{equation}
\begin{array}{l}
O = \sum\limits_{m,n} t^-_m t^-_n H(|\vec r_m - \vec r_n |,\overline E)
( \vec\sigma_m \cdot \vec\sigma_n - 1 )\\
 = - 4 \sum\limits_{m,n} t^-_m t^-_n H(|\vec r_m - \vec r_n |,\overline E)
\hat P_{mn}(S=0)
\end{array}
\end{equation}

\noindent
where $\hat P_{mn}(S=0)$ is the projector of a two-particle state
coupled to spin S=0.
This result will be useful in order to simplify the expressions for the
nuclear matrix element $M_{0\nu}$.

The finite nucleon size and the two-nucleon short range correlations in
the nucleon-nucleon interaction are introduced in these
matrix elements. Their explicit expression are given in Appendix C.

As mentioned above, we will use the pseudo SU(3) scheme to describe the ground
state of the six heavy deformed nuclei considered in this contribution. A
detailed description of this formalism can be found elsewhere
\cite{Cas94,Rat73,Dra84}. In
the next section we give a brief summary of the model.

\vskip 1.0pc

\noindent
\section {The Pseudo SU(3) Model}
\vskip 1.0pc

A tractable shell model theory for deformed nuclei requires a severe truncation
of the spherical model basis. A successful truncation can be achieved if the
basis selection is made relative to those parts of the interaction that
dominate
the low-energy structure: the pairing and quadrupole-quadrupole ones. In the
case of rotational nuclei, it has been shown that the long range  part , {\it
i.e.} the quadrupole-quadrupole interaction, is the most important one
\cite{Dra84}.

For heavy nuclei, the major  shells are built by the orbitals $j = \{ 1/2, \
3/2,
 \dots  \eta - 1/2\}$, which are called of normal parity plus a single particle
state $j = \eta + 3/2$ of a different parity.  Then we are restricting the
Hilbert space to the valence neutron and proton shells constituted by the
corresponding normal parity orbitals and the abnormal or unique parity levels.

In the pseudo $SU(3)$ shell model coupling scheme \cite{Rat73} the normal
parity
orbitals are identified with orbitals of a harmonic oscillator of one quanta
less $\tilde \eta = \eta-1$.  This set of orbitals, with $\tilde j
= j = \tilde  l + \tilde s$,  pseudo spin $\tilde s =1/2$,  and  pseudo angular
momentum  taking  the values $\tilde l =\tilde\eta, \ \tilde\eta -2,
\dots 1 $ or
$0$  define the so called pseudo space and only recently it was found an
analytic expression for the transformation that take us from the normal
parity orbitals to the pseudo space \cite{Cas92}.  Applying this
transformation to the spherical Nilsson hamiltonian it can be
shown explicitly that the strength of the pseudo spin orbit
interaction is almost zero and the orbitals $j = \tilde l \pm
1/2$ are nearly degenerate doublets. For the configurations of
identical particles occupying a single j orbital of abnormal parity, a
convenient characterization of states is made by means of the seniority
coupling scheme.

The many particle states of $n_\alpha$ nucleons in a given shell $\eta_\alpha$,
$\alpha = \nu $ or $\pi$
, can be defined by the totally antisymmetric irreducible representations
$\{ 1^{n^N_\alpha}\} $ and $\{1^{n^A_\alpha}\}$ of unitary groups of dimensions
corresponding to the sizes of the normal  $\Omega^N_\alpha = (\tilde
\eta_\alpha + 1) (\tilde\eta_\alpha +2)$ and unique  $\Omega^A_\alpha =
2\eta_\alpha +4$ parity spaces, respectively; with the constraint
$n_\alpha = n^A_\alpha  + n^N_\alpha$.

 A complete classification of the
states can be defined by the following chains of groups:
\begin{equation}
\begin{array}{ll}
& \matrix{ \
\{ 1^{n^N_\alpha}\} \quad \quad \{ \tilde f_\alpha \} \ \ \qquad
\{f_\alpha \}    \gamma_\alpha   \, (\lambda_\alpha,\mu_\alpha) \quad  \
\tilde S_\alpha   \   \tilde K_\alpha  \ ~~   \tilde L_\alpha
 \qquad  \qquad  \qquad \ \   J^N_\alpha \hfill \\
U(\Omega_\alpha^N) \supset  U(\Omega^N_\alpha/2)
 \times  U(2) \supset  SU(3)  \times  SU(2)  \supset
 SO(3)  \times SU(2)  \supset  SU_J(2) \\} \\
\noalign{\vskip 1truecm} & \qquad \qquad \qquad
\matrix{  \hfill \{1^{n^A_\alpha}\} &
&v_\alpha & \beta_\alpha & J^A_\alpha \\
U(\Omega^A_\alpha) & \supset & Sp (\Omega^A_\alpha) & \supset &
		SU(2) \\} \ \ , \\
\end{array}
\end{equation}

\noindent
where  above each group the quantum numbers that
characterize its irreducible representations (irreps) are given
and the
  $\gamma_\alpha$, $\beta_\alpha$ and $\tilde K_\alpha$ are
multiplicity labels of the indicated reductions. For the normal
parity spaces the pseudo $LS$ coupling scheme is used and the following
relations between its quantum numbers are satisfied
\begin{equation}
n^N_\alpha = f^1_\alpha + f^2_\alpha, \hspace{1cm}
\ S_\alpha = 1/2(f^1_\alpha - f^2_\alpha), \hspace{1cm}
\ \{ \tilde f_\alpha\} = \{2^{n^N_\alpha/2 - S_\alpha}, \ 1^{2S_\alpha}\} \ \ ,
\end{equation}

\noindent
while for
the abnormal parity spaces  the seniority configurations
$v_\alpha$ are appropriate.

Proton and neutron states are coupled in angular momentum in both the
normal and unique parity sectors, generating states with angular momentum
$J^N$ and $J^A$ respectively. The
wave function of the many-particle state  with angular
momentum $J$ and projection $M$ is expressed as a direct product of the
normal  and unique parity ones, as:

\begin{equation}
|J M > = \sum\limits_{J^N J^A} [|J^N> \otimes |J^A>]^J_M
\end{equation}
We will be interested in the ground states of some deformed nuclei, which
have $J=M=0$.

For even-even heavy nuclei, it has been shown that if the
residual neutron-proton interaction is of the quadrupole type,
independently of the interaction in the proton and neutron
spaces, for yrast states below the back bending region the most important
normal parity configurations are
those with highest spatial symmetry $\{ \tilde f_\alpha \} =
\{ 2^{n^N_\alpha /2}\}$. This  implies that $ \tilde S_\pi = \tilde S_\nu = 0$,
that is, only pseudo spin zero configurations are taken into account.

Additionally in the abnormal parity space only seniority zero
configurations,
$v_\pi = v_\nu =0$, are taken into account.  This simplification implies that
$\beta_\pi = \beta_\nu = 1$ and $J^A_\pi = J^A_\nu = 0$. This is a very
strong assumption, which in future works is expected to be improved,
but is quite useful in order to simplify the calculations.

As an example, the ground-state band of  $^{238}U$  is characterized
by the eigenstates
\begin{equation}
\begin{array}{ll}
& | \ \{ 1^6\}_\pi  \{2^3\}_\pi (18,0)_\pi; \ \{
1^{12}\}_\nu \{2^6\}_\nu (36,0)_\nu; \ 1 (54,0) K J M >_N
\\
& | \ (i_{13/2})^4_\pi,  \ J^A_\pi = M^A_\pi = 0; \
(j_{15/2})^8_\nu,
\ J^A_\nu = M^A_\nu = 0 >_A \ \ ,  \\ \label{eq}
\end{array}
\end{equation}

\noindent
where the first term indicates the normal part and the second the
abnormal one. The $^{238}U$ ground state has $K=1, J=M=0$.
 For a detailed description of the way this state was
constructed the reader is referred to \cite{Cas94,Dra84}.

\vskip1.0pc

\section {Calculation of the matrix elements $M_{0 \nu}$}
\vskip 0.5pc

We want to evaluate the nuclear matrix element (4) for several heavy
deformed nuclei, which are potential $ \beta\beta$ emitters.
According to the above discussion the initial $| 0^+_i\rangle$ and final
$| 0^+_f\rangle$ ground states are of the type indicated in the
Eq.(\ref{eq}).

We expand the two body transition operator in order to obtain its normal
and abnormal parity components

\begin{equation}
\begin{array}{l}
O^\alpha = - \sum_{j_1,j_2} \sum_{j_3,j_4} \sum_{J,M}
<(j_1 j_2)JM |O^\alpha_{12} |(j_3 j_4) JM>
[a^\dagger_{j_1} a^\dagger_{j_2} ]^{JM} [a_{j_3} a_{j_4} ]^{JM} \\
= - \sum_{j_\pi,j_{\pi '}} \sum_{j_\nu,j_{\nu '}} \sum_{J,M}
<(j_\pi j_{\pi '})JM |O^\alpha_{12} |(j_\nu j_{\nu '}) JM>
[a^\dagger_{j_\pi} a^\dagger_{j_{\pi '}} ]^{JM}
[a_{j_\nu} a_{j_{\nu '}} ]^{JM}
\end{array}
\end{equation}

\noindent
where in the second line only the nonvanishing contributions were retained.
Transforming this operator to the pseudo $SU(3)$ space, we have
formally the expression
\begin{equation}
O^\alpha =  O^\alpha_{N_\pi N_\nu} +O^\alpha_{N_\pi A_\nu} +
O^\alpha_{A_\pi N_\nu} +O^\alpha_{A_\pi A_\nu}
\end{equation}
\noindent
where the subindices $NN, \ NA, \dots$ are indicating the
normal or abnormal spaces of the fermion creation and annihilation operators,
respectively. By using the Nilsson scheme in order to obtain the
occupation number we are grouping nucleons in pairs. For this reason only
these four cases can give contribution different from zero.

In the present work we will restrict our analysis to six potential
double beta emitters which, within the approximations of the pseudo SU(3)
scheme,   are also decaying via the
$2\nu$ mode. They include the observed \hbox{$^{150}Nd$}$ \rightarrow$
\hbox{$ ^{150}Sm$} and
\hbox{$^{238}U$} $\rightarrow$ \hbox{$ ^{238}Pu$} decays. In this case two
neutrons belonging
to a normal parity orbital decay in two protons belonging to an abnormal
parity one.
Under the seniority zero assumption for nucleons in abnormal
parity orbitals, only proton pairs coupled to J=0 are allowed to exist in
the ground state, restricting the above summation to J=0, M=0.
It follows that $\pi = \pi '$ and $\nu = \nu '$.
The matrix elements becomes
\begin{equation}
\begin{array}{ll}
M_{0\nu}^\alpha =
&- \sum\limits_{j_\pi,j_\nu}
<(j_\pi j_{\pi})J=M=0 |O^\alpha |(j_\nu j_{\nu}) J=M=0> \\
 & < \ \{2^{n^N_\pi/2}\} (\lambda_{\pi}^f,\mu_\pi^f);
\{2^{n^N_\nu/2 -1}\}_\nu (\lambda_{\nu}^f,\mu_\nu^f);
(\lambda^f ,\mu^f) 1 0 0|
[a_{j_\nu} a_{j_{\nu}} ]^{J=M=0} \\
&\hspace{3cm} | \ \{2^{n^N_\pi/2}\} (\lambda_{\pi}^i,\mu_\pi^i);
\{2^{n^N_\nu/2 }\}_\nu (\lambda_{\nu}^i,\mu_\nu^i);
(\lambda^f ,\mu^f) 1 0 0> \\
& < \ (j^A_\pi)^{n^A_\pi+2} ,  \ J^A_\pi = M^A_\pi = 0;
(j^A_\nu)^{n^A_\nu}, \ J^A_\nu = M^A_\nu = 0 |
[a^\dagger_{j_\pi} a^\dagger_{j_{\pi}} ]^{J=M=0} \\
&\hspace{3cm} | \ (j^A_\pi)^{n^A_\pi} ,  \ J^A_\pi = M^A_\pi = 0;
(j^A_\nu)^{n^A_\nu}, \ J^A_\nu = M^A_\nu = 0 >
\end{array}
\end{equation}

In these cases, it is also simple to show that \cite{Cas94}
\begin{equation}
\begin{array}{l}
 < \ (j^A_\pi)^{n^A_\pi+2} ,  \ J^A_\pi = M^A_\pi = 0 \ |
[a^\dagger_{j_\pi} a^\dagger_{j_{\pi '}} ]^{J=M=0}
 | \ (j^A_\pi)^{n^A_\pi} , \ J^A_\pi = M^A_\pi = 0 \ > \\
\hspace{8cm} = \delta_{j^A_\pi j_\pi}
\sqrt{{\frac {(n^A_\pi+2) (2j_\pi+1-n^A_\pi)} {2 j_\pi + 1}}}
\end{array} \label{uniq}
\end{equation}

The operators in the normal parity sector must be transformed to the
pseudo-space. These nucleons are coupled to {\em pseudospin} $\tilde S=0$,
which together with the J=0 condition implies $\tilde L = 0$. Then
\cite{Cas94,Dra84}
\begin{equation}
[a_{j_\nu} a_{j_\nu}]^{J=M=0} = \sqrt{\frac {2 j_\nu +1}
{2(2\tilde l_\nu+ 1)}}
[\tilde a_{(0,\tilde \eta_\nu) \tilde l_\nu ;
{\frac 1 2}} \tilde a_{(0,\tilde \eta_\nu) \tilde l_\nu ; {\frac 1 2}}]^{
\tilde L = \tilde S = 0 ; J = M = 0}
\end{equation}

Expanding the normal states in their proton and neutron subspaces, and the
above operator in their SU(3) components, together with Eq.
(\ref{uniq}),  we arrive to the final expression

\begin{equation}
\begin{array}{ll}
M_{0\nu}^\alpha &\equiv \sum\limits_{j_\nu} M_{0\nu}^\alpha (j_\nu ) \\
&= -\sum\limits_{j_\nu} \sqrt{\frac {2 j_\nu +1} {2(2\tilde l_\nu + 1)}}
\sqrt{{\frac {(n^A_\pi+2)(2j_\pi +1-n^A_\pi)} {2 j_\pi + 1}}}
<(j^A_\pi j^A_\pi )J=M=0 |O^\alpha |(j_\nu j_\nu ) J=M=0>  \\
&\sum\limits_{K_\pi L_\pi} \sum\limits_{K_\nu K'_\nu}
<(\lambda_\pi^f,\mu_\pi^f)K_\pi L_\pi ,
(\lambda_\nu^f,\mu_\nu^f)K'_\nu L_\pi \|(\lambda^f,\mu^f) 1 0 >  \\
&\hspace{2cm}<(\lambda_\pi^i,\mu_\pi^i)K_\pi L_\pi ,
(\lambda_\nu^i,\mu_\nu^i)K_\nu L_\pi \|(\lambda^i,\mu^i) 1 0 >  \\
&\sum\limits_{(\lambda_0 ,\mu_0 ) \rho_0}
<(\lambda_\nu^i,\mu_\nu^i)K_\nu L_\nu , (\lambda_0 ,\mu_0 ) 1 0 \|
(\lambda_\nu^f,\mu_\nu^f) K'_\nu L_\nu >_{\rho_0} \\
&\hspace{2cm}< (0,\tilde \eta_\nu) 1 \tilde l_\nu , (0,\tilde \eta_\nu) 1
\tilde l_\nu \| (\lambda_0 ,\mu_0 ) 1 0> \\
&\hspace{2cm}<(\lambda_\nu^f,\mu_\nu^f) \||
[\tilde a_{(0,\tilde \eta_\nu) \tilde l_\nu ; {\frac 1 2}}
\tilde a_{(0,\tilde \eta_\nu) \tilde l_\nu ; {\frac 1 2}}]^{
(\lambda_0 ,\mu_0 );\tilde S = 0} \||
(\lambda_\nu^i,\mu_\nu^i)>_{\rho_0} \label{m0nu}
\end{array}
\end{equation}
\noindent
where the $<(\lambda_1,\mu_1)K_1 L_1 , (\lambda_2 ,\mu_2 ) K_2 L_2 \|
(\lambda ,\mu ) K L >_{\rho}$ are SU(3) Clebsch-Gordan coefficients, and
$<(\lambda_1,\mu_1)\||O\||(\lambda_2,\mu_2)>$ are triple reduced matrix
elements \cite{Cas94}. We have implicitly defined
$M_{0\nu}^\alpha (j_\nu) $ as the contribution of each normal
parity neutron state $j_\nu$ to the nuclear matrix element in the transition
$(j_\nu )^2 \rightarrow (j_\pi^A)^2$.

The two-body matrix element can be expanded in its $L,S$ components

\begin{equation}
\begin{array}{l}
<(j_\pi j_\pi )J=M=0 |O^\alpha |(j_\nu j_\nu ) J=M=0> \\
\hspace{3cm}= \sum_S \sqrt{(2 j_\pi + 1) (2 j_\nu + 1)} (2S+1)
W({\frac 1 2} j_\pi S l_\pi ; l_\pi {\frac 1 2})
W({\frac 1 2} j_\nu S l_\nu ; l_\nu {\frac 1 2})\\
\hspace{5cm}<(l_\pi l_\pi )S\| H(r, \overline E) \| (l_\nu l_\nu) S>
<({\frac 1 2}{\frac 1 2}) S \| \Gamma \cdot \Gamma (\alpha) \|
({\frac 1 2}{\frac 1 2}) S >
\end{array}
\end{equation}

where the W(...;...) are  Racah coefficients and

\begin{equation}
\Gamma \cdot \Gamma (GT) \equiv \vec\sigma_1 \cdot \vec\sigma_2
\hspace{1cm}
\Gamma \cdot \Gamma (F) \equiv 1
\hspace{1cm}
\alpha = GT ~\hbox{or}~ F   .
\end{equation}

In order to evaluate the spatial matrix elements, we use the
Bessel-Fourier expansion of the potential \cite{Hor61}, which gives

\begin{equation}
\begin{array}{l}
<(l_1 l_2 ) L M | V(r) | (l_3 l_4 ) L M > \\
\hspace{2cm}= \sum_l (-1)^{l_1 + l_4 + L} (l_1\|C_l\|l_3) (l_2\|C_l\|l_4)
W(l_1 l_2 l_3 l_4;L l) (2l+1) R^l(l_1 l_2,l_3l_4)
\end{array}
\end{equation}

\noindent
with $(l_1\|C_l\|l_3)$ the reduced matrix elements of the unnormalized
spherical harmonics
$ C_{lm}(\Omega ) \equiv \sqrt{\frac {4\pi}{2l+1}} Y_{lm}(\Omega )$
and $R^l(l_1 l_2 , l_3 l_4)$ the radial integrals which are explained in
Appendix A.

Although we will present results for $M_{0\nu}^F$ and $M_{0\nu}^{GT}$, we
have shown that using $\frac {g_A^2} {g_V^2} = 1.0$ only the $S=0$
channel survive, {\em i.e.} only states with L=0 are of
interest and, after some manipulations, we obtain the simplified expression

\begin{equation}
\begin{array}{l}
<(j_\pi j_\pi )J=M=0 | O |(j_\nu j_\nu ) J=M=0> \\
\hspace{4cm}= -2 \sqrt {\frac {(2 j_\pi+1)(2 j_\nu+1)}{(2 l_\pi+1)(2
l_\nu+1)}}
<(l_\pi l_\pi ) L=0 | H(r,\overline E) | (l_\nu l_\nu ) L=0  >\\
\hspace{4cm} = -2 \sqrt { (2 j_\pi+1)(2 j_\nu+1) }
\sum\limits_l (-1)^l ( l_\nu 0 l_\pi 0 | l 0)^2
R^l(l_\pi l_\pi ,l_\nu l_\nu )    .
\end{array}
\end{equation}

The radial integrals $R^l(l_\pi l_\pi ,l_\nu l_\nu )$
are evaluated in the way described in Appendix A,
and they include finite nucleon size and short range correlations, as
explained in Appendix C.

\vskip 1.0pc

\noindent
\section{Results and discussion}
\vskip 1.0pc
We evaluated the matrix elements of $M_{0\nu}$ between the ground states of
six different initial and final nuclei for
the $ \beta\beta$ transitions. There exists a relevant dependence of this
transition amplitudes with the partition $\{n^A_\pi , n^N_\pi , n^A_\nu
, n^N_\nu \}$ of the
valence nucleons of each pair of nuclei. We used the standard procedure to fix
them, taking deformation from experiment, and filling each Nilsson level with a
pair of particles in order of increasing energies \cite{Cas94,Dra84}. As
mentioned in the Introduction, we restricted the present analysis to
nuclei in which the unique and normal proton and neutron valence
occupations are  related
by
\begin{equation}
\begin{array}{l}
n^A_{\pi ,f} = n^A_{\pi ,i} + 2,
\hspace{1cm}n^A_{\nu ,f} = n^A_{\nu ,i} \\
n^N_{\pi ,f} = n^N_{\pi ,i} ,
\hspace{1.7cm} n^N_{\nu ,f} = n^N_{\nu ,i} - 2
\end{array}
\end{equation}

\noindent
The six nuclei fulfilling these conditions are listed in Table 1, and are
the same for which finite half-lifes were predicted in the two neutrino
case \cite{Cas94}. Their matrix elements $M_{0\nu}$ are shown in the second
column.
 The value for the integrated kinematical factor,
$G_{0 \nu}$ in [years$^{-1}$] was obtained following the procedure
indicated
by Doi et al \cite{Doi88} with $g_A/g_V = 1.0$ and it is given in the third
column. In the fourth
 column of Table 1, the energy value $\overline E$ is indicated and in the
last two
columns the theoretical predictions ( for $<m_e>=1eV$) and experimental
lower limits of the $\beta\beta_{0\nu}$-half lives are given.
It has been demonstrated that these $M_{0\nu}$ matrix elements are nearly
insensitive to the exact values of the average energies if they are in
the range of some MeV \cite{Pan92}. We choose the accepted parametrization
$\overline E = 1.12 A^{1/2}$ \cite{Hax84}.

As mentioned above, for $^{238}U$ the predicted $0\nu$ half live
$\tau_{0\nu}^{1/2} = 1.0 \times 10^{24}$ years is three orders of
magnitude greater than the predicted $2\nu$ half live
$\tau_{0\nu}^{1/2} = 1.4 \times 10^{21}$ years, which essentially agrees
with the experimental one, confirming that the observed $\beta\beta$
decay of $^{238}U$ has to be the two neutrino mode. Figure 1 exhibits the
$M_{0\nu}^{GT}(j_\nu )$ and $-M_{0\nu}^F(j_\nu )$ components of the
Gamow-Teller and Fermi nuclear matrix elements for $^{238}U$. It can be
seen that all the terms contribute constructively, and that the
transition is strongly dominated by a pair of neutrons in the normal
parity orbital $i_{11/2}^\nu$
decaying into two protons in the unique orbital $i_{13/2}^\pi$,
resembling  the two neutrino case (see \cite{Cas94} and Appendix B).

In the case of $^{150}Nd$, the pseudo SU(3) $0\nu$ matrix element
reported here is about a factor four lesser than the QRPA estimations.
This is a very relevant result. First, it exhibits the stability of the
neutrinoless double beta decay matrix elements evaluated in quite
different nuclear models, in the case of deformed nuclei. Second, this
factor of four, which is little compared with the order of magnitude
variations in the $2\nu$ theoretical estimations, is still important in
order to extract the parameters $<m_e>$ and $<g_{\nu ,\chi}>$.
In Figure 2 the
$M_{0\nu}^{GT}(j_\nu )$ and $-M_{0\nu}^F(j_\nu )$ components of the
Gamow-Teller and Fermi nuclear matrix elements for $^{150}Nd$ are shown.
As for $^{238}U$ all the terms contribute constructively, but in this case
the transition is  dominated by a pair of neutrons in the normal
parity orbital $h_{9/2}^\nu$
decaying into two protons in the unique orbital $h_{11/2}^\pi$, again
resembling  the two neutrino case.

As can be seen in the last two columns of Table 1, the
$\tau_{0\nu}^{1/2}$ predicted for $<m_\nu > = 1 eV$ are at least three
order of magnitude greater than the experimental limits. These results
reflect the fact that at, the present stage of the experimental
$\beta\beta$ research, the limits $<m_\nu > \le 1.1 eV$ obtained by the
Heidelberg-Moscow collaboration \cite{Mai94} using significative volumes
of ultrapure $^{76}Ge$ are the most sensitives. But, if the
$\beta\beta_{0\nu}$ decay is
observed in $^{76}Ge$, at least a second observation will be essential,
and $^{150}Nd$ is a likely candidate to do this job \cite{Moe94}.
In the next few years the limit for $<m_\nu >$ extracted from
$\beta\beta_{0\nu}$ experiments is expected to be improved up to 0.1 eV,
and $^{150}Nd$ is one of the selected isotopes \cite{Nem94}.

On the other side, in extracting the majoron coupling constant the
$^{150}Nd$ has set the lowest limit \cite{Moe94}. The phase integral for
the two emitted electrons and the majoron, with ${\frac {g_A^2} {g_V^2}}
= 1.0$, for this nuclei is
\begin{equation}
G_{0\nu \chi} = 0.466 \times 10^{-14} yrs^{-1}
\end{equation}

\noindent
and the reported lower limit for this decay is \cite{Moe94}
\begin{equation}
\tau^{1/2}_{0\nu \chi} > 5.30 \times 10^{20} yrs   .
\end{equation}

Using these numbers in Eq. (3) we obtain the limit
\begin{equation}
<g_{\nu ,\chi}> \le 4.1 \times 10^{-4}  .
\end{equation}

This limit is less stringent than the previous one \cite{Moe94} obtained
with the same experimental information but employing the QRPA matrix elements
$M_{0\nu}^{QRPA} = 6.075$ \cite{Mut89}.

There are various elements which contributes to
the greater cancellation of the pseudo SU(3) $M_{0\nu}$ matrix elements,
as compared with the QRPA ones.

i) The uncertainty in the QRPA results itself: the other QRPA
reported
value for $^{150}Nd$ is $M_{0\nu}=5.24$ \cite{Tom91}, $14\%$ lesser than
the above mentioned one.

ii) The way short range correlation effects were parametrized in
$H(r,\overline E)$ in the present work is not the most common one. But both
finite nucleon size (FNS) and two-nucleon short range correlations (SRC)
effects together reduce the $0\nu$ matrix elements in less than $15\%$.

iii) Another source of cancellation
 could be related with the seniority zero
approximation used to describe nucleons in abnormal parity states.
Removing this assumption in Eq. (\ref{m0nu}) would lead to a more general
expression, not restricted to states with $J = M = 0$. Being far from
trivial, this extension of the pseudo SU(3) model is under investigation.
But as can be seen in Fig. 3 of \cite{Mut89}, the $(j_\pi j_\pi ) J=0;
(j_\nu j_\nu ) J=0$ channel represents by far the most relevant
contribution to the $0\nu$ matrix elements, and inclusion of other
multipoles could {\em decrease} $M_{0\nu}$.

iv) Only one active shell was allowed for protons, and one for neutrons.
This is a very strong truncation, of the same type as used in shell model
calculations. Extensions of the pseudo SU(3) model which takes fully into
account core excitations have shown that the SU(3) component of the wave
function are the dominant ones, between 50\% and 60\% of the total
amplitude \cite{Cas92a}.
Inclusion of these symplectic excitations could increase the present
results, allowing for example transitions from neutron states to proton
states {\em in the same shell}. Given the SU(3) dominance, these
corrections are estimated to contribute at most with a factor two.

Nevertheless, at the present stage of development, the pseudo SU(3)
predictions are as valid as the QRPA ones, which suffer other kind of
limitations, and their differences are in some way describing the
theoretical uncertainties in the extraction of the parameters $<m_\nu>$
and $<g_{\nu ,\chi}>$.

\vskip 1.0pc

\noindent
\section{Conclusions}
\vskip 1.0pc

Our description of the neutrinoless $\beta\beta$ decay in heavy deformed
nuclei can be summarized as follows:

The pseudo SU(3) model is a very powerful machinery to describe the
collective behavior of heavy deformed nuclei. Together with
a model hamiltonian with spherical Nilsson single particle terms,
and a residual interaction of a quadrupole-quadrupole type, plus terms
including some scalar combinations of the quadrupole and orbital angular
momentum operators, it has been used to reproduce very accurately the
rotational spectra of heavy deformed nuclei, including the K-band splitting,
the amplitudes for transitions of the E2, M1 and M3 type and the $2 \nu
\beta\beta$ decay of eleven heavy deformed nuclei, with good agreement
with the available experimental data.

The deformed ground states were described using the pseudo SU(3) scheme,
which is able to generate maximum deformation for nucleons
in normal parity orbitals, filling them into the Nilsson levels but with
definite
total angular momentum. These states are strongly correlated, and are
dominated by the proton-neutron quadrupole-quadrupole interaction,
considered the microscopic origin of deformation.  The nucleons in the
abnormal parity  orbitals are
assumed to have seniority zero, i.e. in this approximation they do not
contribute actively to the quadrupole moments.
This is the strongest approximation of the model, only
justified in terms of its simplicity.

We have evaluated the $\beta\beta_{0\nu}$ half lives of six heavy
deformed nuclei, using the pseudo SU(3) approach together with the closure
approximation, obtaining an order of magnitude agreement with
the QRPA estimations, and a factor four of difference. We
exhibited predictions for the neutrinoless double beta half-lives
assuming $<m_\nu>=1 eV$, and discussed the relevance of our results.
In the case of $^{238}U$ this result complements those obtained for the two
neutrino $\beta\beta$ decay, and it confirms the observed half-life as two
neutrino in origin.
Limits for the majoron coupling constant were extracted from the $^{150}Nd$
experimental limit, and compared with the QRPA ones.

\newpage
\appendix{\bf Appendix A: The radial integrals}

\vskip 1cm
The radial integrals are defined as\cite{Hor61}

\begin{equation}
\begin{array}{l}
R^l(l_1 l_2 , l_3 l_4) \equiv \int\limits_0^\infty v(p) p^2 \\
\hspace{3cm}\{ \int\limits_0^\infty r_1^2 dr_1 j_l(pr_1)
R_{n_1 l_1}(r_1) R_{n_2 l_2}(r_2)
\int\limits_0^\infty r_2^2 dr_2 j_l(pr_2)  R_{n_3 l_3}(r_1) R_{n_4
l_4}(r_2) \} dp \\  \\
\hspace{2cm} v(p) = {\frac 2 \pi} \int\limits_0^\infty H(r) j_0(pr) r^2 dr
\end{array}
\end{equation}

\noindent
The $R_{n l}$ are the one-particle radial harmonic oscillator wave
functions and $j_0(pr)$ is the 0-th order spherical Bessel function.

Using the explicit form of $H(r,\overline E)$, Eq. (1), we find \cite{Krm92}
\begin{equation}
v(p) = {\frac {2R} {\pi}} {\frac 1 {p(p+\overline E)}}
\end{equation}

The explicit expression for the radial integral is \cite{Hor61}

\begin{equation}
\begin{array}{l}
R^l(l_1 l_2 , l_3 l_4) = ( M(n_1 l_1,n_3 l_3) M(n_2 l_2,n_4 l_4) )^{-1/2} \\
\sum\limits_{m1,m2,m} a_{m_1}(n_1 l_1,n_3 l_3) a_{m_1}(n_2 l_2,n_4 l_4)
a_{2m}(\frac{m_1- l}2 l ,\frac{m_2-l}2 l )  J_m
\end{array}
\end{equation}
\noindent
where $m_1 = l_1 + l_3 + 2 s$ and $s = 0,1,...,n_1+n_3$, and similarly to
$m_2$,  $m = l, l+1,...,\frac{m_1+m_2}2 $,

\begin{equation}
\begin{array}{l}
M(n l, n' l') = 2^{n+n'} n! n'! (2l+2n+1)!! (2l'+2n'+1)!!\\
a_{l+l'+2s}(nl, n'l') = (-1)^s \sum\limits_{\mu, \mu'=s-\mu}
{\frac{n!}{\mu ! (n-\mu)!}}{\frac{n'!}{\mu' ! (n'-\mu')!}}
{\frac {(2l+2n+1)!!}{2l+2\mu +1)!!}}{\frac {(2l'+2n'+1)!!}{2l'+2\mu' +1)!!}}
\end{array}
\end{equation}
\noindent and
\begin{equation}
J_m = (b^2/2)^m \int\limits_0^\infty exp({\frac{-p^2 b^2}2})v(p)p^{2m+2}dp
\end{equation}
being $b=\sqrt{\frac{\hbar}{m\omega}}$ the oscillator length.

In the case $\overline E = 0$, we obtain
\begin{equation}
\begin{array}{l}
H(r,\overline E) = H(r,0) = \frac R r\\
v(p)= {\frac{2R}{\pi p^2}}  \hspace{2cm}J_m = {\frac{\sqrt{2} R}{\pi b}}
\Gamma (m+1/2)
\end{array}
\end{equation}

\noindent
which is a not so bad approximation to the exact result \cite{Pan92},
obtained using
a nonzero $\overline E$ and including the short range correlations and
finite nucleon size effects, as explained in Appendix C.

\vskip 3cm

\appendix{\bf Appendix B: The two-neutrino limit}

\vskip 1cm
It is interesting to exhibit how the two neutrino matrix elements
presented in \cite{Cas94} can be obtained from the present formalism.
In the two neutrino case $H(r) \rightarrow 1$, and the transition
operators are the double Fermi or the double Gamow-Teller ones.
In this limit
\begin{equation}
\int_0^\infty j_l(pr_1) j_l(pr_2) v(p) p^2 dp =
\frac 1 2 \int_{-1}^1 H(r) P_l(cos \omega) d(cos \omega )
\rightarrow \delta_{l 0}
\end{equation}
\noindent
where $r = |\vec r_1 - \vec r_2|$ and $\omega$ is the angle between $\vec
r_1$ and $\vec r_2$ and
\begin{equation}
\begin{array}{ll}
R^l(l_1 l_2 ,l_3 l_4 ) &\rightarrow \delta_{l 0}
\delta_{n_1 n_3} \delta_{n_2 n_4} \\
<(l_1 l_2) LM | V(r) | (l_3 l_4) LM> &\rightarrow
\delta_{n_1 n_3} \delta_{l_1 l_3} \delta_{n_2 n_4} \delta_{l_2 l_4}
\end{array}
\end{equation}

\noindent where $n_i$  are the number of nodes in the radial
wave function $R_{n_i l_i}$.

Using the previous results and expressing $j_\pi = l_\pi +1/2, j_\nu =
l_\pi - 1/2$ we obtain
\begin{equation}
\begin{array}{l}
<(j_\pi j_\pi )J=0 | O | (j_\nu j_\nu ) J=0 > \rightarrow
\delta_{n_\pi n_\nu} \delta_{l_\pi l_\nu} 2 \sqrt{l_\pi(l_\pi +2)}\\
\hspace{2cm} \sum\limits_S (2S+1) W({\frac 1 2} l_\pi+{\frac 1 2} S l_\pi
; l_\pi {\frac 1 2}) W({\frac 1 2} l_\pi-{\frac 1 2} S l_\pi ; l_\pi
{\frac 1 2}) <({\frac 1 2} {\frac 1 2}) S \| \Gamma \cdot \Gamma \|
({\frac 1 2} {\frac 1 2}) S > \\
= \delta_{n_\pi n_\nu} \delta_{l_\pi l_\nu}
 {\sqrt{\frac {l_\pi (l_\pi +1)}{2l_\pi + 1} }}
 \{ <({\frac 1 2} {\frac 1 2}) S=0 \| \Gamma \cdot \Gamma \|
({\frac 1 2} {\frac 1 2}) S=0 > -
<({\frac 1 2} {\frac 1 2}) S=1 \| \Gamma \cdot \Gamma \|
({\frac 1 2} {\frac 1 2}) S=1 >  \}  .
\end{array}
\end{equation}

In the case of double Fermi transitions, {\em i.e.}
$\Gamma \cdot \Gamma = \sum_{m n} t^-_m t^-_n$, the above described matrix
elements are always
zero. This result was argued heuristically in \cite{Cas94}, and exhibit
the isospin as a good quantum number for the pseudo SU(3) ground state.
For double Gamow-Teller transitions, where $\Gamma \cdot \Gamma =
\sum_{m n} \vec\sigma_m \cdot \vec\sigma_n t^-_m t^-_n$ we obtain
\begin{equation}
<(j_\pi j_\pi )J=0 | O^{GT} | (j_\nu j_\nu ) J=0 > \rightarrow
- \delta_{n_\pi n_\nu} \delta_{l_\pi l_\nu} 4
{\sqrt{\frac {l_\pi (l_\pi +1)}{2l_\pi +1}}}\\
\end{equation}

\noindent which substituted in Eq. (\ref{m0nu}) gives the $<0^+_f|\Gamma
\cdot \Gamma |0^+_i>$ in exact equivalence with Eq. (6.8) and (6.9) of
\cite{Cas94}.
This $2\nu$ double Gamow-Teller operator creates a pair of protons coupled to
total angular
momentum zero  and annihilates two neutrons of the  normal parity space
 coupled to pseudo orbital angular momentum and
 pseudo spin equal to zero. For the rare earth  and actinide nuclei this
operator annihilates two neutrons in the pseudo shells  $\eta_\nu = 5$ and
$\eta_\nu = 6$ and creates two protons in the abnormal orbits $h_{11/2}$ and
$i_{13/2}$, respectively.

\vskip 2cm
\appendix{\bf Appendix C:  The finite nucleon size and short range
correlation effects}

\vskip 1cm
In the momentum representation it is not difficult to include additional
effects.
The finite nucleon size (FNS) effects are introduced through the dipole form
factor \cite{Tom91,Tom87}

\begin{equation}
v_{FNS}(p) = v(p) \{ {\frac {\Lambda^2}{\Lambda^2+p^2}} \}^4
\hspace{2cm} \Lambda = 850 MeV
\end{equation}

The two-nucleon short range correlations (SRC) mainly arise from the
repulsion due to the $\omega$-exchange in the nucleon-nucleon
interaction \cite{Bro77}. Thus the correlated potential is constructed as

\begin{equation}
\begin{array}{l}
v_{SRC}(p) = \int {\frac {d^3k} {(2\pi )^3}} \Omega(\vec p - \vec k ) v(p)\\
\Omega(\vec p - \vec k ) \equiv (2\pi )^3 \delta (\vec p) -
{\frac {2\pi^2} {q_c^2}} \delta (q - q_c)
\end{array}
\end{equation}

\noindent with $q_c = 3.93 fm^{-1}$ roughly the Compton wavelength of the
$\omega$ meson. Within the approximation $q_c \ll \overline E$ we get

\begin{equation}
v_{SRC}(p) = v(p) - \Delta v(p)
\hspace{2cm}
\Delta v(p) \equiv {\frac {2\pi} {q q_c}} ln |{\frac {q+q_c} {q-q_c}}|
\end{equation}

When both the FNS and the SRC effects are considered the neutrino
potential takes the form:

\begin{equation}
v_{FNS+SRC}(p) = v_{FNS} - \Delta v(p) + \Delta v'(p)
\end{equation}

\noindent with

\begin{equation}
\Delta v'(p) \equiv {\frac \pi {p q_c}} \{\sum_{n=1}^3 \frac 1 n
[x_-^n - x_+^n] + ln {\frac {x_-^n} {x_+^n}} \}
\hspace{1cm}
x_\pm^n \equiv {\frac {\Lambda^2} {\Lambda^2 + (p \pm q_c )^2}}  .
\end{equation}

It was shown in Ref. \cite{Krm92} that the FNS effects are negligible for
the allowed transitions but increases rapidly when L increases, and that,
if both FNS and SRC are considered simultaneously the net result may not
be very different from that obtained without any SRC at all.

\vfill
\eject

\vfill
\eject
\vskip 2cm
\centerline{{\bf  Table Captions}}
\vskip 1.0truecm
\noindent
{\bf Table 1.}  Theoretical estimates for the nuclear matrix
elements and the half-life of the
$\beta\beta$-decay in the $0\nu$ mode for several heavy deformed
nuclei are given, under the assumption $<m_e> = 1 eV$, and compared with the
available experimental data.

\vskip 2cm

\centerline{{\bf Table 1}}

$$
\begin{array}{crcccc}
\hline
\\
\hbox{Transition}  &M_{0\nu} &G_{0\nu}[yrs^{-1}] &\overline E [MeV]
&\tau^{1/2}_{theo} &\tau^{1/2}_{exp} \nonumber \\ \\
\hline
\\
^{146}Nd \to ^{146}Sm &1.16 &1.65 \times 10^{-17} &13.5
&1.18 \times 10^{28} & \nonumber \\ \\
\hline
\\
^{148}Nd \to ^{148}Sm &1.57 &1.57 \times 10^{-14} &13.6
&6.75 \times 10^{24} \nonumber \\ \\
\hline
\\
^{150}Nd \to ^{150}Sm &1.57 &1.01 \times 10^{-13} &13.7
&1.05 \times 10^{24} &> 2.1 \times 10^{21} \, \cite{Moe94} \nonumber \\  \\
\hline
\\
^{186}W \to ^{186}Os &1.70 &1.76 \times 10^{-15} &15.3
&5.13 \times 10^{25} &> 2.3 \times 10^{20} \, \cite{Dan93}\nonumber \\ \\
\hline
\\
^{192}Os \to ^{192}Pt &0.72 &1.58 \times 10^{-15} &15.5
&3.28 \times 10^{26}\nonumber  \\ \\
\hline
\\
^{238}U \to ^{238}Pu &1.76 &8.21 \times 10^{-14} &17.3 &1.03
\times 10^{24} &> 2.0 \times 10^{21} \, \cite{Tur91} \nonumber \\
\nonumber \\ \hline \nonumber
\end{array}
\nonumber
$$

\newpage
\vskip 2cm
\centerline{Figure Captions}
\vskip 2cm
Figure 1. The $M_{0\nu}^{GT}(j_\nu )$ (white columns) and
$-M_{0\nu}^F(j_\nu )$ (black columns) components of
the Gamow-Teller and Fermi nuclear matrix elements for $^{238}U$, as
function of the angular momentum of the decaying neutrons.

\vskip 1cm
Figure 2. The same of Fig. 1 for $^{150}Nd$.
\vfill
\newpage
\begin{figure}
\leavevmode
\epsfxsize=5in
\caption{}
\end{figure}
\newpage
\begin{figure}
\leavevmode
\epsfxsize=5in
\caption{}
\end{figure}

\end{document}